\documentclass{iau}
\usepackage{graphicx}
\usepackage{hyperref}

\title[JD 11.~~Tidal evolution in multiple planet systems] 
{Tidal evolution in multiple planet systems: application to Kepler-62 and Kepler-186}

\author[E. Bolmont, S. N. Raymond, J. Leconte, A. Correia \& E. Quintana]   
{Emeline Bolmont$^{1,2}$, Sean N. Raymond$^{1,2}$, J\'er\'emy Leconte$^{3,4,5}$, Alexandre Correia$^{6,7}$ \and Elisa Quintana$^{8,9}$}

\affiliation{$^1$Univ. Bordeaux, Laboratoire d'Astrophysique de Bordeaux, UMR 5804, F-33270, Floirac, France \\[\affilskip]
$^2$CNRS, Laboratoire d'Astrophysique de Bordeaux, UMR 5804, F-33270, Floirac, France\\[\affilskip]
$^{3}$Canadian Institute for Theoretical Astrophysics, 60st St George Street, University of Toronto, Toronto, ON, M5S3H8, Canada\\[\affilskip]
$^{4}$Banting Fellow\\[\affilskip]
$^{5}$ Center for Planetary Sciences, Department of Physical \& Environmental Sciences,\\
 University of Toronto Scarborough, Toronto, ON, M1C 1A4\\[\affilskip]
$^6$Departamento de F'sica, I3N, Universidade de Aveiro, Campus de Santiago, 3810-193 Aveiro, Portugal\\[\affilskip]
$^7$ASD, IMCCE-CNRS UMR8028, Observatoire de Paris, UPMC, 77 Av. Denfert-Rochereau, 75014 Paris, France\\[\affilskip]
$^8$SETI Institute, 189 Bernardo Ave, Suite 100, Mountain View, CA 94043, USA\\[\affilskip]
$^9$NASA Ames Research Center, Moffett Field, CA 94035}

\pubyear{2014}
\volume{310}  
\pagerange{xxx--xxx}
\jname{Complex Planetary Systems}
\editors{Z. Knezevic \& A. Lema\^itre}
\begin{document}

\maketitle

\begin{abstract}
A large number of observed exoplanets are part of multiple planet systems. Most of these systems are sufficiently close-in to be tidally evolving. In such systems, there is a competition between the excitation caused by planet-planet interactions and tidal damping. Using as an example two multiple planet systems, which host planets in the surface liquid water habitable zone (HZ): Kepler-62 and Kepler-186, we show the importance and effect of both planetary and stellar tides on the dynamical evolution of planets and on the climate of the HZ planets. 
\keywords{Planets and satellites: dynamical evolution and stability, Planet-star interactions,  stars: individual (Kepler-62, Kepler-186), methods: N-body simulations.}
\end{abstract}

\firstsection 
\section{Introduction}

More than 1400 exoplanets have now been detected and about 20\% of them are part of multiple planet systems (\url{http://exoplanets.org/}). Most of these planetary systems have close-in planets for which tides have an influence. In particular, tides can have an effect on the eccentricities of planets, and also on their rotation periods and their obliquities, which are important parameters for any climate studies (so-called Milankovitch cycles; \cite[Berger, 1988]{berger1988}). Besides tides can influence the stability of multiple planet systems, by their effect on the eccentricities but also by their effect on precession rates.

We present here a few examples of tidal evolution of multiple planet systems including the study of the Kepler-62 (\cite[Borucki et al. 2013]{Borucki2013}) and Kepler-186 (\cite[Quintana et al. 2014]{Quintana2014}) systems. These recently discovered systems host 5 planets. Kepler-62 e and f are in the HZ of a $0.69~M_\star$ star. Kepler-186 f is an Earth-sized planet in the HZ of a $0.42~R_\star$ star. We study the possible evolution of these planets' orbits as well as their spin states. 

\section{Method}

In order to study the dynamical evolution of these systems, we used a new code, which takes into account the evolution of the radius and spin of the central object (be it a star or brown dwarf evolving according to the evolution tracks of \cite[Chabrier et al. 2000]{Chabrier2000}). It also includes a rigorous treatment of tidal forces (\cite[Hut, 1981]{Hut1981}; \cite[Leconte et al. 2010]{Leconte2010}), the effect of the rotation-induced flattening of planets and star (\cite[Correia et al. 2013]{Correia2013}), the correction due to general relativity (\cite[Kidder, 1995]{Kidder1995}), and planet-planet gravitational interactions using a symplectic N-body algorithm (\cite[Chambers, 1999]{Chambers1999}). This code will be introduced in Bolmont et al. 2015 (to be submitted). 

We used as initial orbital elements, stellar and planetary parameters the values given by \cite[Borucki et al. (2013)]{Borucki2013} for Kepler-62 and \cite[Bolmont et al. (2014)]{Bolmont2014} for Kepler-186. In \cite[Bolmont et al. (2014)]{Bolmont2014}, we derived two sets of planetary parameters: set A and set B both consistent with the data and self-consistent. These two sets were used here. 

We derived the masses of the planets for different planetary compositions using \cite[Fortney et al. (2007)]{Fortney2007}. We chose for the Kepler-62 system an Earth-like composition for all planets. For Kepler-186, we tested compositions from 100\% ice to 100\% iron. 

\section{Results}

\subsection{Stability}

We find that tides influence the stability of multiple planetary systems. This is especially clear in the case of Kepler-62 for which simulations without general relativity and tides lead to a destabilization in a few $10^7$~yr. However simulations with general relativity and tides are stable during the 30 million years of the simulation. We also find that the system is very sensitive to the chosen planets' masses: a change of a few percent can stabilize significantly the system. 

When planetary tides are taken into account, the stability of the system depends on the chosen planets' tidal dissipation. Instabilities tend to occur when the planets' dissipation are high. This is slightly counter-intuitive because increasing dissipation should contribute to decrease the eccentricity and stabilize the system. Changing dissipation changes the frequencies of the system, and some configurations happen to lead to a destabilization. This can be used as a way of reducing the size of the parameter space: dissipations leading to unstable configurations are unlikely. However, this could change with the tidal model used. Here we use the constant time lag model (\cite[Hut, 1981]{Hut1981}) and we plan to investigate how these features change using other tidal models.

A more thorough study of the stability of this system is needed, based on more longer simulations or different methods (such as \cite[Laskar, 1990]{Laskar1990}, \cite[Correia et al. 2005]{Correia2005}, \cite[Couetdic et al. 2010]{Couetdic2010}), and should take into account all the effects considered here.

\subsection{Obliquity and rotation period evolution}

For Kepler-62, the rotation period of the three inner planets of the system evolves towards pseudo-synchronization in less than ten million years and their obliquities evolve towards small equilibrium values ($<1^\circ$). Given as the age of the system is estimated at 7 Gyr (\cite[Borucki et al. 2013]{Borucki2013}), we expect that the three inner planets of the Kepler-62 system are now slowly rotating (their period is higher than 100 hr) and they have quasi null obliquities. 

\begin{figure}[htbp]
\begin{center}
 \includegraphics[width=\linewidth]{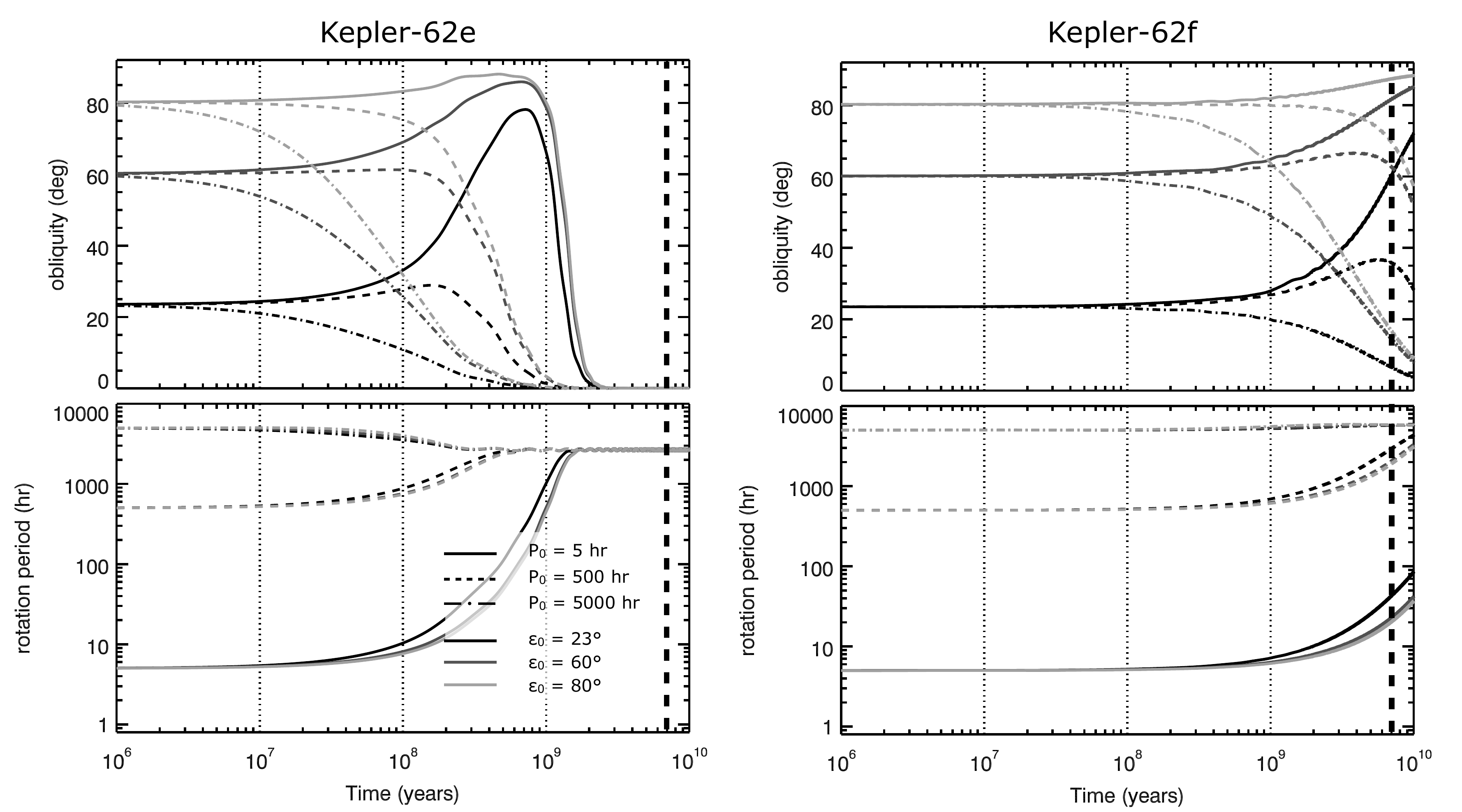} 
 \caption{Evolution of the obliquity and the rotation period evolution for Kepler-62e and f for different initial obliquities (23$^\circ$, 60$^\circ$ and 80$^\circ$) and different initial rotation periods (5 hr, 500 hr and 5000 hr). The thick vertical black line corresponds to the estimated age of the system: 7 Gyr (\cite[Borucki et al. 2013]{Borucki2013}).}
   \label{Kepler_62_obl_rot}
\end{center}
\end{figure}

During the 30 million years of the simulation, the obliquities and rotation periods of the HZ planets Kepler-62e and f do not evolve significantly. So we performed longer simulations for these two outer planets. 
Assuming an Earth-like dissipation for the two planets, we found that Kepler-62e is likely to have reached pseudo-synchronization and have low obliquity (see Figure \ref{Kepler_62_obl_rot}). The dissipation of the planets is not constrained and changing the dissipation would only shift the curves right (if the dissipation is lower) or left (if the dissipation is higher). As an Earth-like dissipation is actually probably a high dissipation value (\cite[Lambeck, 1977]{Lambeck1977}), it is likely that the curves should be shifted to the right. For Kepler-62f the timescales of evolution are higher and Figure \ref{Kepler_62_obl_rot} shows that the rotation period is still evolving towards pseudo-synchronization after 7~Gyr of evolution and that the obliquity can still be high. 

Likewise, for Kepler-186, the four inner planets are located close-in to the star (orbital periods less than 25 days) and their rotation rates evolve in less than a million year. Given that the system is thought to be older than 4 Gyr (\cite[Quintana et al. 2014]{Quintana2014}), we expect these four inner planets to be pseudo-synchronized and have low obliquities (\cite[Bolmont et al. 2014]{Bolmont2014}).

However, the tidal evolution timescales of the HZ planet Kepler-186f (orbital period of 130 days) are longer. Long term simulations showed us that giving the high uncertainty on the dissipation of this planet, it might still be tidally evolving or it might have reached the tidal equilibrium. If the dissipation of Kepler-186f is low or if the planetary system is young ($\sim$1~Gyr), the planet might have a high obliquity (up to 80$^\circ$) and a relatively fast rotation ($\sim$30~hr). However, if the dissipation of Kepler-186f is high or the system is old (more than a few Gyr), the planet might have a low obliquity and a slow rotation ($\sim$3000~hr).



\subsection{An extra planet in the system Kepler-186}

In \cite[Bolmont et al. 2014]{Bolmont2014}, we show that formation simulations tend to form more than the 5 detected planets in a system like Kepler-186. We thus proposed the idea that there could be a sixth undetected planet in the system in the gap between planet e and f. However, this extra planet is not in a transit configuration because it would have been detected otherwise. We performed simulations to see if the dynamics of a system with an extra planet could be stable and consistent with the observations. We tested this hypothesis for different masses for the extra planet: from 0.1~$M_\oplus$ to $M_{{\rm Jupiter}}$, with an initial inclination of 2$^\circ$. 

Without an extra planet, Kepler-186f is dynamically isolated. Adding an extra planet in the system causes the inclination of the orbit of Kepler-186f to be excited. The more massive the planet, the higher the excitation of the inclination. If the mass of the extra planet is more than 1~$M_\oplus$, the inclination of Kepler-186f oscillates with values higher than the detectability limit (i$_{{\rm lim}}$=arctan($R_\star$/$a$), where $R_\star$ is the radius of the star and $a$ the semi-major axis of the planet). Thus, if an extra planet does exist, it should have a relatively low mass: $\lesssim 1~M_\oplus$.

\section{Conclusions}


We showed here the importance of tides for the evolution of the planetary systems Kepler-62 and Kepler-186. 

Taking into account tides increases the stability of the Kepler-62 system unless we assume the planets to be very dissipative. 


We investigated the evolution of the spin states of Kepler-62e, Kepler-62f and Kepler-186f. The climate of an exoplanet depends strongly on its orbital elements and also its rotation. If it has a low obliquity and near synchronous rotation, the planet might have a cold trap (Kepler-62e). If it has a high obliquity and rapid rotation, it would have strong seasonal effects and Coriolis induced wind patterns (Kepler-62f). Kepler-186f could be in either configurations, the age of the system and the tidal dissipation are very poorly constrained.   

Finally, we showed that there could be an extra planet in the Kepler-186 system between planets e and f. If such a planet existed, it would be a low mass planet $\lesssim 1~M_\oplus$.

%


\begin{thebibliography}{}

\bibitem[Berger, 1988]{berger1988}
{Berger, A.} 1988, 
\textit{Reviews of Geophysics}, 26, 624

\bibitem[Bolmont et al. 2014]{Bolmont2014}
{Bolmont, E., Raymond, S. N., von Paris, P., et al.} 2014,
\textit{ApJ}, 793, 3

\bibitem[Borucki et al. 2013]{Borucki2013}
{Borucki, W. J., Agol, E., Fressin, F., et al.} 2013,
\textit{Science}, 340, 587

\bibitem[Chabrier et al. 2000]{Chabrier2000}
{Chabrier, G. \& Baraffe, I.} 2000, 
\textit{ARAA}, 38, 337

\bibitem[Chambers, 1999]{Chambers1999}
{Chambers, J. E.} 1999, 
\textit{MNRAS}, 304, 793

\bibitem[Correia et al. 2005]{Correia2005}
{Correia, A. C. M., Udry, S., Mayor, M., et al.} 2005, 
\textit{A\&A}, 440, 751

\bibitem[Correia et al. 2013]{Correia2013}
{Correia, A. C. M., \& Rodr\'iguez, A.} 2013, 
\textit{ApJ}, 767, 128

\bibitem[Couetdic et al. 2010]{Couetdic2010}
{Couetdic, J., Laskar, J., Correia, A. C. M., Mayor, M., \& Udry, S.} 2010, 
\textit{A\&A}, 519, A10

\bibitem[Fortney et al. 2007]{Fortney2007}
{Fortney, J. J., Marley, M. S., \& Barnes, J. W.} 2007, 
\textit{ApJ}, 659, 1661

\bibitem[Hut, 1981]{Hut1981}
{Hut,P.} 1981,
\textit{A\&A}, 99,126

\bibitem[Kidder, 1995]{Kidder1995}
{Kidder, L. E.} 1995, 
\textit{Phys. Rev. D}, 52, 821

\bibitem[Lambeck, 1977]{Lambeck1977}
{Lambeck, K.} 1977, 
\textit{Royal Society of London Philosophical Transactions Series A}, 287, 545

\bibitem[Laskar, 1990]{Laskar1990}
{Laskar, J.} 1990,
\textit{Icarus}, 88, 266

\bibitem[Leconte et al. 2010]{Leconte2010}
{Leconte, J., Chabrier, G., Baraffe, I., \& Levrard, B.} 2010, 
\textit{A\&A}, 516, A64+

\bibitem[Quintana et al. 2014]{Quintana2014}
{Quintana, E. V., Barclay, T., Raymond, S. N., et al.} 2014,
\textit{Science}, 344, 277


\end{thebibliography}
\end{document}